\documentclass[aps,prc,showpacs]{revtex4-1}                  
\usepackage{graphicx}
\usepackage{amsmath} 
\usepackage{amssymb} 
\usepackage{graphics}
\usepackage{epsfig}
\usepackage{slashed}
\newcommand\ba{\begin{eqnarray}}
\newcommand\ea{\end{eqnarray}}
\newcommand\be{\begin{equation}}
\newcommand\ee{\end{equation}}

\newcommand{\bas}{\begin{eqnarray*}}
\newcommand{\eas}{\end{eqnarray*}}

\begin{document}

\title{Two-photon exchange: myth and history} 


\author{Egle Tomasi-Gustafsson }      \email[Corresponding author:]{egle.tomasi@cea.fr}
\affiliation{IRFU, CEA, Universit\'e Paris-Saclay, 91191 Gif-sur-Yvette, France }
  
\author{
        Simone Pacetti
}
\affiliation{Dipartimento di Fisica e Geologia, and INFN Sezione di Perugia, 06123 Perugia, Italy}

\begin{abstract}
After recalling the arguments for possible excess of two-photon contribution over $\alpha$-counting,
 model independent statements about the consequences on the observables will be given.
The relevant experimental data are discussed:  (polarized and unpolarized) electron and positron elastic scattering on the proton, as well as  annihilation data. A reanalysis of unpolarized electron-proton elastic scattering data is presented in terms of the electric to magnetic form factor squared ratio. This observable is in principle more robust against experimental correlations and global normalizations. The present analysis shows indeed that it is a useful quantity that contains reliable and coherent information. The comparison with the  ratio extracted from the measurement of the longitudinal to transverse polarization of the recoil proton in polarized electron-proton scattering shows that the results are compatible. These results bring a decisive piece of information in the controversy on the deviation of the proton form factors from the dipole dependence.
\end{abstract}

\maketitle
\section{Introduction}
\label{intro}
Hadron electromagnetic form factors (FFs) describe the internal structure of hadrons in terms of their electric and magnetic charge distributions. They  constitute a very convenient parametrization of the hadron electromagnetic current on the basis of a very well established formalism that assumes the exchange of a virtual photon of four-momentum $Q^2$ ($1\gamma$ E). In this framework, FFs are a privileged background for theory and experiments. They are directly accessible through differential cross section and polarization observables of elementary reactions: elastic electron(positron)-proton scattering, $e^\pm p\to e^\pm p$, and  the crossed reactions, the annihilations $e^++e^- \leftrightarrow \bar p + p $. 
The {\it reduced} cross section of  electron-proton elastic scattering, in the Born approximation, i.e., by considering only one-photon exchange,  $\sigma_{\rm red}$,  is linear in the variable $\epsilon=[1+2(1+\tau)\tan^2(\theta_e/2)]^{-1}$, being $\theta_e$ the electron scattering angle in the proton rest frame.  This formalism assumes that the exchange of two virtual photons ($2\gamma E$) is small. Assuming simple $\alpha$-counting, ($\alpha=e^2/4\pi=1/137)$, the interference of $1\gamma E$ and $2\gamma E$ would not exceed $1\%$ of the total amplitude. The two photon contribution itself would be of the order of  $0.01\%$ to the cross section. The possibility to evidence the presence of $2\gamma E$ is therefore related to the fact that the interference gives rise to charge-odd observables, as charge asymmetry in electron versus positron elastic scattering on protons. In the annihilation region a forward-backward asymmetry would be observed. One can prove, in model independent way, that the presence of  odd $\cos\theta$ terms (where $\theta$ is the emission angle of one final particle in the center of mass system) has a kinematical correspondence  with the $\epsilon$ linearity in the Rosenbluth plot  \cite{Rekalo:1999mt}. Moreover,  non vanishing single spin polarization in unpolarized $ep$ elastic scattering would also be a signature of $2\gamma E$. 

The two photon contribution was widely discussed in the literature in the 70's~\cite{Gunion:1972bj,Boitsov:1972if,Franco:1973uq,Lev:1975}, and its experimental evidence was searched for. No conclusive result was found, in the limit of the experimental precision. As a conclusion of a series of measurements (for a review, see \cite{TomasiGustafsson:2009pw}), no experimental evidence was found. Since that time, the $1\gamma E$ approximation was assumed {\it a priori}. The  two ($n$)- photon contribution may be observed only if other mechanisms compensate the factor $Z\alpha$ $((Z\alpha)^n)$ that scales the size of the amplitude. One reason for which $2\gamma E$  may become important at large transferred momentum is that, if the transferred momentum is equally shared between the two photons, the steep decreasing of FFs (calculated for $Q^2/2$) may compensate the scaling in $\alpha$. In this context, it is expected that $2\gamma E$ becomes more important when $Q^2$ increases and/or when the charge $Z$ of the target increases. Moreover, as the $2\gamma E$ amplitude contains, in principle, an imaginary part, it could be enhanced in the time-like region, where FFs are complex. 

The presence of a sizable  $2\gamma E$ contribution was more recently reproposed for $ed$ elastic scattering \cite{Rekalo:1999mt}, to explain discrepancies between two experiments on elastic electron-deuteron scattering at Jefferson Laboratory (JLab) \cite{Alexa:1998fe,Abbott:1998sp}. The differences in the cross sections, at similar $Q^2$ values, but at different incident beam energies and electron scattering angles, were not increasing with $Q^2$, suggesting instead a systematic shift of the hall C spectrometer position (a shift of $0.3^\circ$ of the central angle was indeed found). 

Very recently,  expensive and extensive experimental and theoretical work was focussed  to the search of evidence of  $2\gamma E$, due to the suggestion that it could explain the discrepancy between the unpolarized and polarized measurements of FFs from $ep$ elastic scattering  \cite{Guichon:2003qm}. The favored method to measure elastic FFs was based to the 'Rosenbluth separation'  \cite{Rosenbluth:1950yq}: the measurement of the unpolarized cross section for a fixed $Q^2$ at different angles. It turns out that this method is limited by the precision on the extraction of the electric FF, at large $Q^2$, as the magnetic contribution is enhanced by a factor of $\tau=Q^2/4M^2$, $M$ being the proton mass. 
The possibility of very precise measurements at large transferred  momentum was opened by  the development of 100\% duty cycle electron machines as JLab, with highly polarized electron beams, the construction of large solid angle spectrometers and detectors. The development of proton polarimetry in the GeV region made possible to apply the polarization method suggested by A.I. Akhiezer and M.P. Rekalo at the end of the sixties \cite{Akhiezer:1968ek,Akhiezer:1974em}, stimulating large experimental and theoretical work devoted to hadron FFs. These authors pointed out that the polarization transferred from a longitudinally polarized electron beam to a polarized proton target (or the measurement of the polarization of the recoil proton) in elastic electron proton scattering contains a term of interference between the electric and magnetic amplitudes, being more sensitive to a small electric contribution, and also to its sign. The FF ratio, $G_E/G_M$, is proportional to the ratio of the longitudinal to transverse polarization of the recoil proton, $P_L/P_T$.

The data on the FF ratio, collected mostly by the GEp collaboration at JLab (\cite{Puckett:2017flj} and References therein) show that, not only the precision is larger as expected, but also that the ratio, normalized to the proton magnetic moment, deviates from unity, as previously commonly accepted. Meaningful data were collected up to $Q^2\simeq$ 9 GeV$^2$.  A discrepancy, increasing with $Q^2$, appeared between polarized and unpolarized elastic scattering experiments, giving rise to a large number of publications and speculations. In particular, the comparison was focussed on the  work from Ref. ~\cite{Andivahis:1994rq}, as it extends the individual FF extraction to the largest values of  $Q^2$, and on a recent dedicated experiment at JLab \cite{Qattan:2004ht}, both based on the Rosenbluth method.

The purpose of this work is to revise critically the claimed evidence of $2\gamma E$, on the basis of model independent statements and existing data. The electron probe has been considered a very clean way to access the hadron structure, as far as radiative corrections are kept under control. If $2\gamma E$   becomes dominant at large $Q^2$, a serious revision of most of the physics accessible through electron scattering would be needed. Some information on the nucleon content can still be derived, may be, but at the price of a very complicated formalism and the measurement of additional observables. 

\section{Recent literature}

In Ref. \cite{Guichon:2003qm}, the presence of  $2\gamma E$  was proposed as a solution of the discrepancy for the proton FFs, the physical reason being  '{\it an accidental amplification}'. Indeed, such contribution induces a more complicated structure of the reaction amplitudes, therefore a larger flexibility in describing the data. The suggested parametrization \cite{Guichon:2003qm} was based on the ansatz that the  $2\gamma E$ amplitude would be real and linear in the $\epsilon $ variable. Let us note that these hypotheses contradict the nature of $2\gamma E$. 

A series of articles on model independent properties of the two-photon contribution on different processes:
$ep$ scattering~\cite{Rekalo:2003xa,Rekalo:2003km,Rekalo:2004wa},
$\bar p p \to e^+e^-$~\cite{Gakh:2005wa}, $ e^+e^-\to \bar p p$~\cite{Gakh:2005hh,Chen:2008hka,Zhou:2011yz},
 showed that the hadronic current is  parametrized  by three structure functions, of complex nature and depending on two  kinematical variables, instead that by two FFs functions of $Q^2$ (that are real in the space-like region). $2\gamma E$ induces, in principle, non-linearities in the Rosenbluth fit as the amplitudes depend explicitly on $\theta$, not only on $Q^2$.  The extraction of the real  Sachs FFs would still be possible, but requiring: either polarized electron and positron beams applying the Akhiezer-Rekalo method to the sum of the cross sections (where odd terms disappear), or measuring five T-even or three T-odd polarization observables, including triple spin observables, which appears very difficult as they are expected to be of the order of $\alpha$.
 
  \begin{figure}
  \begin{minipage}[c]{0.65\textwidth}
    \includegraphics[width=\textwidth, trim = 14mm 42mm 15mm 23mm, clip ]{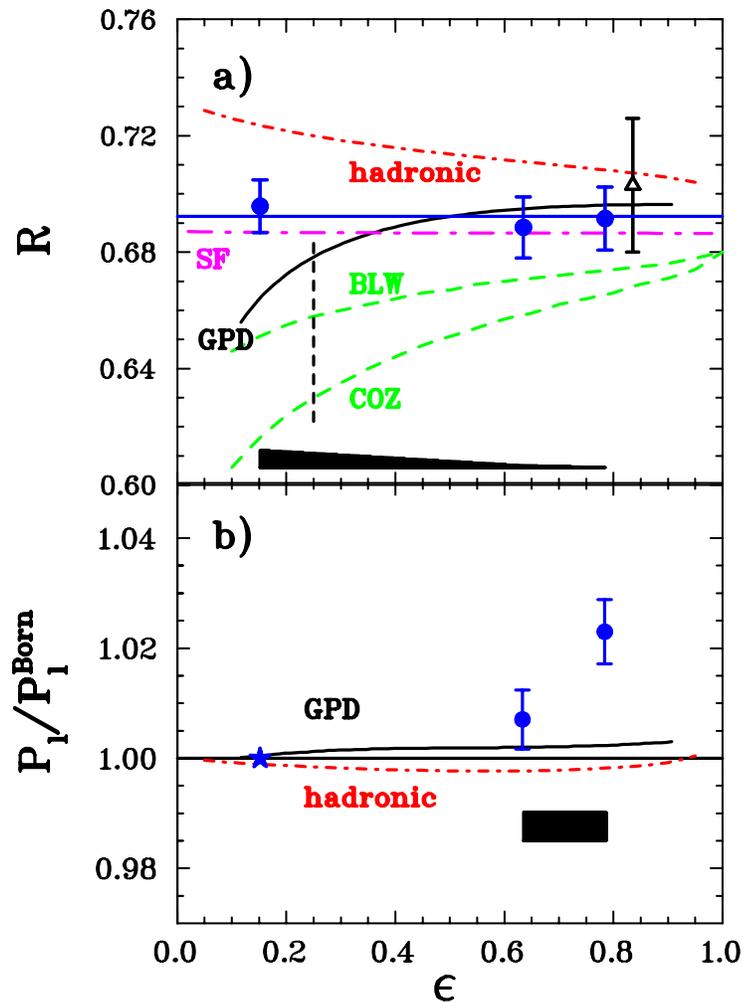}
  \end{minipage}
   \hspace*{-1true cm}
  \begin{minipage}[c]{0.4\textwidth}
 \caption{From Ref. \cite{Meziane:2010xc}. a) $R=G_E/G_M$ as a function of $\epsilon$ with statistical uncertainties, filled circles from this experiment and open triangle 
  from Ref. \cite{Punjabi:2005wq}. The theoretical predictions are from Refs \cite{Blunden:2005ew}, 
  \cite{Afanasev:2005mp}, \cite{Kivel:2009eg} and \cite{Bystritskiy:2007hw} (offset for clarity by -0.006 with respect to the fit. b) $P_{\ell}/P^{Born}_{\ell}$ as a function of $\epsilon$.  }
 \label{Fig:ratiopl} 
  \end{minipage}
\end{figure}

A larger effect of $2\gamma E$ is expected for heavier hadrons, as the expansion parameter is  
$Z\alpha$. A model independent analysis of $e^-  He^4$ scattering and $e^+ e^-\to \pi^+ \pi^-$~was done in Ref. \cite{Gakh:2008fb} not showing the need to introduce contributions beyond $1\gamma E$.  Effects induced by (odd number) multi photon exchange were studied in Ref. \cite {Kuraev:2009hj} (and References therein),  suggesting to detect forward $e^+$ and $e^-$ scattering as well as $p$ and $\bar p$ scattering on heavy target: a universal correction to the Rutherford cross section, which could be experimentally observable, was derived. 

Reanalyses of $e^+p/e^-p$ data~\cite{Arrington:2007ux,TomasiGustafsson:2009pw,Alberico:2009yp}, searching for non-linearities of the reduced cross section,  gave no evidence of charge asymmetry.  

\begin{figure}
  \begin{center}
    \includegraphics[width=0.9\columnwidth]{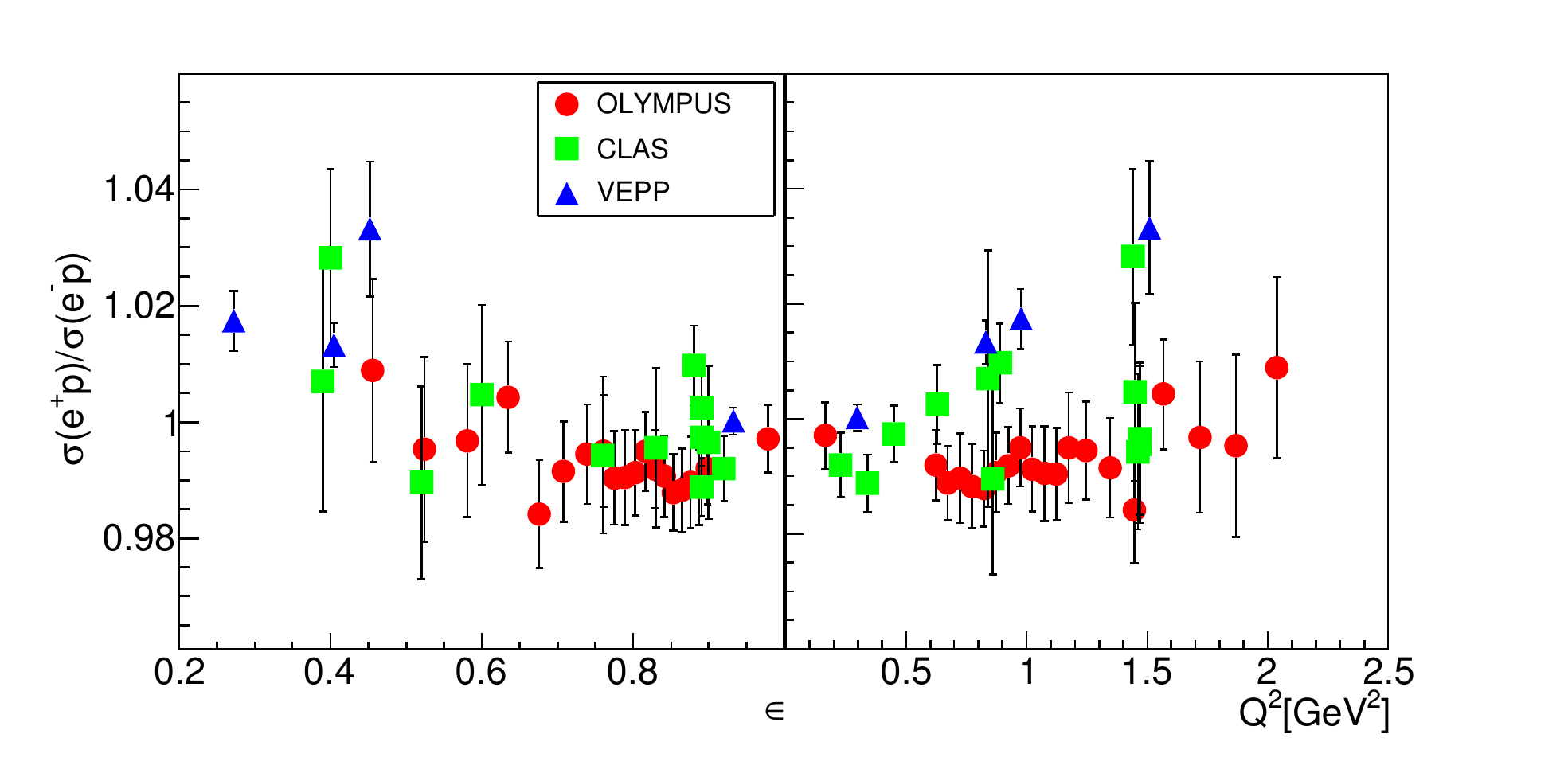}
  \caption{Radiatively corrected ratio of positron to electron cross sections $R=\sigma(e^+p)/ \sigma(e^-p)$, as a function of $\epsilon$ (left) and $Q^2$ (right) from Olympus \cite{Henderson:2016dea} (red circles), CLAS \cite{Rimal:2016toz} (green squares)  and VEPP-3 \cite{Rachek:2014fam} (blue triangles).  
}
 \label{Fig:Alldata} 
    \end{center}
\end{figure}

Several model calculations of the hadronic $2\gamma E$ contribution appeared, with different quantitative results since the numerical calculations, as well as the physical reasons for an enhancement of this term beyond the $\alpha$-counting expectation, differ essentially from a model to another~\cite{Afanasev:2005mp,Borisyuk:2008es,Kivel:2009eg,Blunden:2005ew}. We do not enter here in the comparison and the merit of the existing model dependent $2\gamma E$ calculations. Let us note that, if a qualitative agreement may be found on reproducing the difference between polarized and unpolarized FF ratio, the agreement disappears when compared to another observable, the $\epsilon$ dependence of $P_L/P_T$, as shown in Fig.  \ref{Fig:ratiopl} taken from Ref. \cite{Meziane:2010xc}.  Only the calculation \cite{Bystritskiy:2007hw}, based on high order radiative corrections obtained with the structure function method 
\cite{Kuraev:1988xn,Kuraev:1985hb}, reproduces the results both on the unpolarized cross section and on the polarization ratio. This is due to the fact that $\epsilon$ non linearities from this calculation are very small. This calculation shows also that radiative corrections increase with $Q^2$ and induce a large $\epsilon $ and $Q^2$ dependence  in the individual  longitudinal and transverse polarized cross sections, what explains the deviation of $P_L$ from the Born expectation (Fig.  \ref{Fig:ratiopl}b),  although they essentially cancel in  the ratio, Fig. \ref{Fig:ratiopl}a. 
New measurements were proposed at VEPP-3, Novosibirsk \cite{Rachek:2014fam,Nikolenko:2015xsa},  at CLAS (JLab) \cite{Rimal:2016toz}, and at Olympus (DESY) \cite{Henderson:2016dea}. The results show that  an asymmetry between electron and positron scattering exists indeed, and may reach 6-7\%, but most of the asymmetry comes from the interference between initial and final photon emission, and is highly reduced when the data are properly radiatively corrected. The size of additional $2\gamma E $ contribution does not exceed  the expected size from $\alpha$-counting (few \%), see Fig. \ref{Fig:Alldata}. The main conclusion of these works is that  (difficult) measurements at larger $Q^2$ are necessary: the present results are performed at  $Q^2 \le 3$ GeV$^2$ and do not show  evident increase with $Q^2$. A coherent increase is seen in the most precise VEPP data, that depend, however, on a normalization between two data sets at different beam energy. Note that an effect growing with $Q^2$ and reaching $6\%$ is necessary to bring in agreement the data on the ratio $G_E/G_M$, based on the  Akhiezer-Rekalo and the Rosenbluth methods.

Several other issues concerning the FFs discrepancy were discussed in the literature: radiative corrections were revisited \cite{Bystritskiy:2007hw,Gramolin:2016hjt,Gerasimov:2015aoa}, correlations between the parameters were recalculated \cite{TomasiGustafsson:2006pa}, relative normalization within a set of data and among sets of data \cite{Pacetti:2016tqi,Arrington:2003df}. 

Here we focus on the reanalysis  of Ref. \cite{Pacetti:2016tqi}, concerning in particular the data from Ref. \cite{Andivahis:1994rq}. A careful reading of that work lead us to the conclusion that the discrepancy among this set and the GEP data may be induced by an arbitrary renormalization.

\section{Analysis of unpolarized $ep$ elastic scattering data}

In Ref. \cite{Pacetti:2016tqi} it was suggested to write the reduced cross section for unpolarized $ep$ elastic scattering in terms of the FF ratio $R=G_E(Q^2)/G_M(Q^2) $ : 
\be 
\sigma_{\rm red}(Q^2,\epsilon)= G_M^2(Q^2)[R^2(Q^2) \epsilon +\tau]\,, 
\label{eq:sred}
\ee
where $G_M^2$ and $R^2$ are the independent parameters of a linear fit of the cross section as a function of $\epsilon$ at a fixed $Q^2$. If such procedure is equivalent in most cases to the more usual extraction of $G_E$ and $G_M$ taken as independent parameters, it may differ when $G_E$ is very small and/or in case of large correlations. The ratio is directly extracted, by automatically accounting for the effect of the correlations between $G_E$  and $G_M$. The parameter $R^2$ represents directly the deviation of the linear dependence of the cross section from a constant term in $\epsilon$, whereas general normalization and systematic errors would be absorbed by $G_M^2$. 

The data of Ref.~\cite{Andivahis:1994rq} are especially interesting, with eight $Q^2$ points and  two spectrometer settings, called 8 GeV and 1.6 GeV, spanning the region $1.75 \le  Q^2 \le 8.83$~GeV$^2$.  The two settings will be indicate as  high energy (HE) and low energy (LE) experiments. 

In the original paper the measured cross sections were published, warning that an uncertainty of $\pm 5\%$ affected the second setting, due to a poor knowledge of the acceptance of the spectrometer. This error, however, was not added to the tabulated error on the cross section, but it was introduced just for the FF extraction.  It was taken into account as a constant relative correction. A specific procedure was applied, following three steps. 
\begin{enumerate}
\item  For the two lowest values $Q^2 =1.75$ and $2.50 $ GeV$^2$, the cross section was measured at both settings for the lowest $\epsilon$. It showed a 4-5\% larger value for the LE setting. 
\item The linear $\epsilon$ dependence of the reduced cross section, i.e., the dominance of $1\gamma E$ was assumed.  A linear fit of the HE data was done and the LE energy point was then renormalized to sit on the line. 
\item The same constant normalization $C=0.956$, fixed on the low $Q^2$ point,  was then applied to all measurements taken with the LE setting.
\end{enumerate}
This procedure has the effect to enhance the slope, increasing the FF ratio. Note that for $Q^2=6$ and 7~GeV$^2$ only two points are present. The renormalization (lowering) of the first point changes completely the slope of the linear fit (Analysis I).

\begin{figure}
  \begin{minipage}[c]{0.55\textwidth}
       \includegraphics[width=\textwidth, trim = 4mm 2mm 5mm 3mm, clip ]{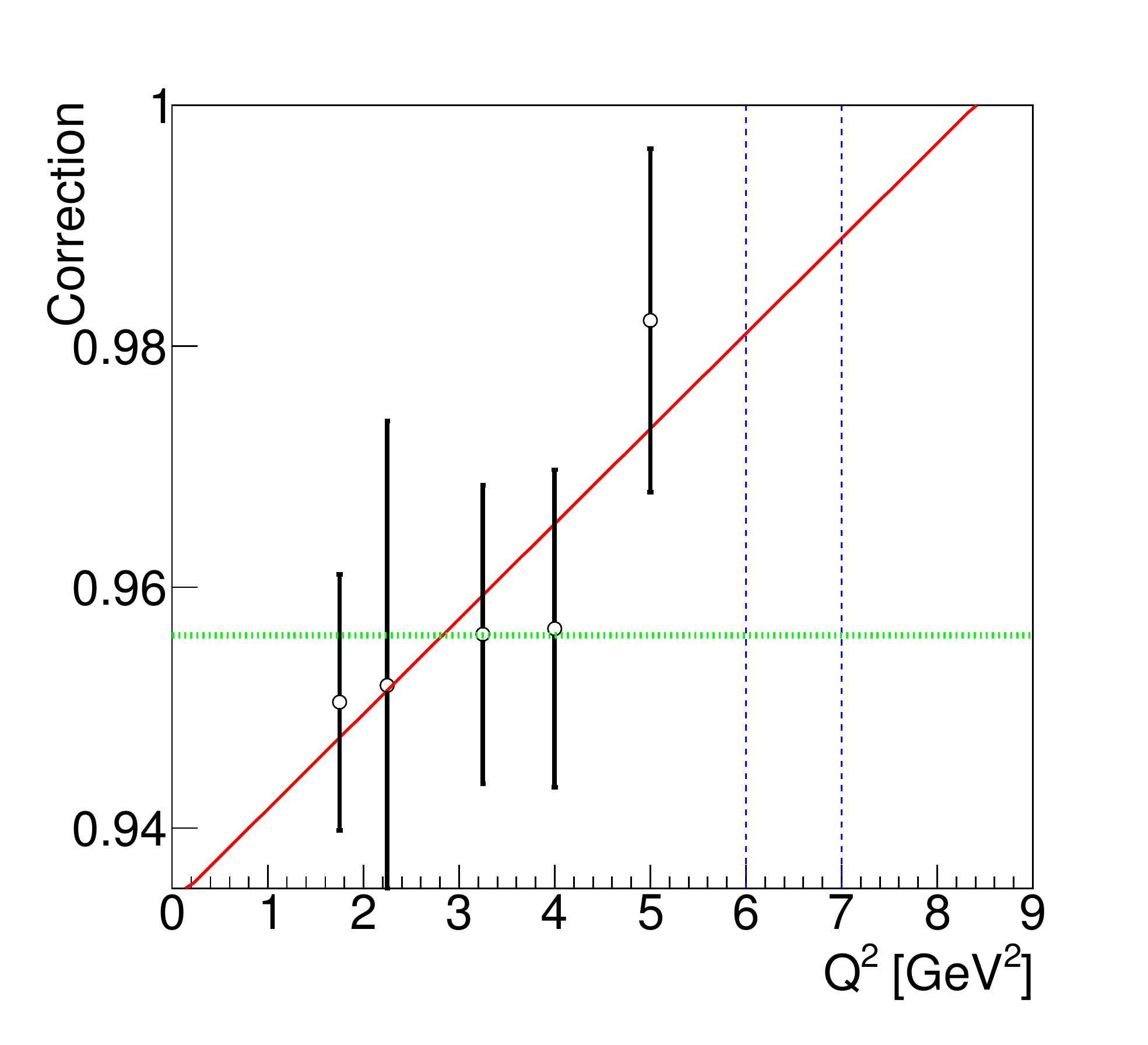}
  \end{minipage}
  \begin{minipage}[c]{0.45\textwidth}
\caption{Correction factor as a function of $Q^2$. A linear fit (red line) shows an increasing of the factor. The dashed (blue) lines indicate that the extrapolated correction for the two HE points would be close to $1\%$ instead than $\simeq 5\%$, as applied in the original paper.}
\label{Fig:correction}
 \end{minipage}
\end{figure}

\begin{figure}
  \begin{minipage}[c]{0.5\textwidth}
    \includegraphics[width=\textwidth, trim = 4mm 2mm 5mm 3mm, clip ]{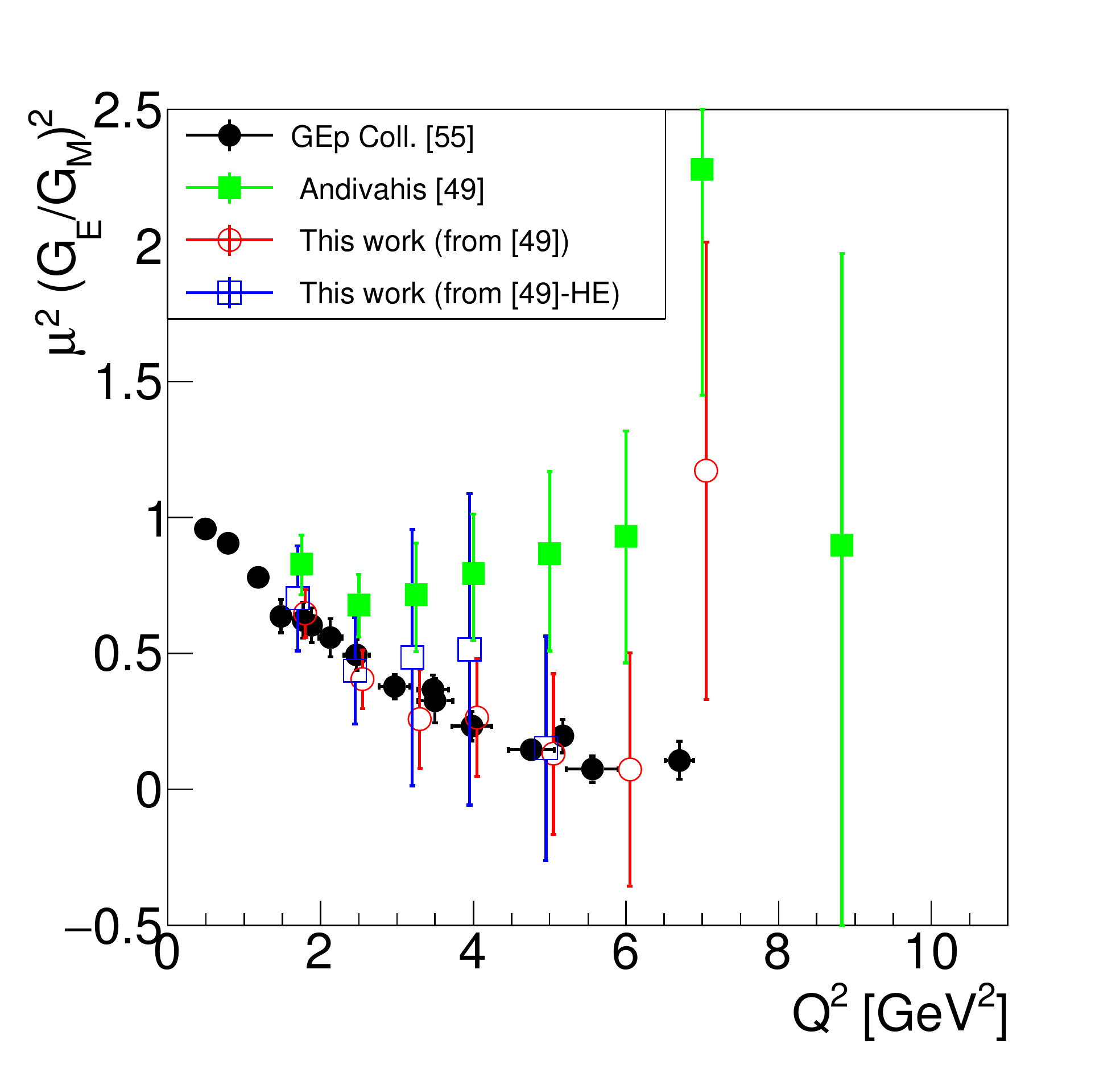}
 \end{minipage}
  \begin{minipage}[c]{0.5\textwidth}
\caption{$\mu^2 R^2= \mu^2 (G_E/G_M)^2$  as a function of $Q^2$ from Ref.~\cite{Andivahis:1994rq} as originally published (green solid squares); from Analysis II: without renormalization (red open circles); from Analysis III: omitting the lowest $\epsilon$ point (blue open  squares)  compared to the values from polarization experiments~\cite{Puckett:2011xg} (black solid circles). 
 }
\label{Fig:W0}
 \end{minipage}
\end{figure}

If the systematic error affecting the low energy point  is related to the acceptance of the setup, in principle it should not be constant with the particle momenta. Therefore, we tried other extractions of the FFs, from the published cross section data. Analysis II: We  recalculated the ratio using the data as published, without renormalizing the two settings and considering the LE points as additional, independent measurements. The data points at $Q^2 =1.75 $ and $2.5 $~GeV$^2$ were both included in the fit, constraining the fit to an average value. Analysis III: the LE points were ignored and only the HE points were fitted (excluding therefore the points at  $Q^2=6$ and 7~GeV$^2$). In the last case we found a slope consistent with Analysis II, although affected by larger errors, as the number of points is smaller.  Analysis IV: we repeated the normalization procedure, by aligning the LE point on the straight line fitting the HE points. We noted a systematic increase of the normalization factor  (Fig. \ref{Fig:correction} and Table \ref{table1}).  We found that the needed corrections decrease at large energies ($C \to 1$). If we apply a normalization coefficient that is derived from the condition that the LE point sits on the straight line fitted on the HE points gives the same slope and intercept as for Analysis III. This explains the agreement between Analyses III and IV. 
\begin{table}[h!]
\begin{center}
\begin{tabular}{|c|c|}
\hline\hline
 $ Q^2$ (GeV$^2$)& Correction \\
\hline\hline
1.75$^*$ &  0.951144 $\pm$ 0.0156952 \\
1.75         &  0.950432 $\pm$  0.0106094 \\
2.25$^*$  &  0.955992 $\pm$  0.0259077 \\
2.25         &  0.951849 $\pm$  0.0219368 \\
3.25         &  0.956075 $\pm$  0.0123809  \\
4              &  0.956552 $\pm$  0.0131748 \\
5              &  0.982138 $\pm$  0.0142443 \\                         
 \hline \hline
\end{tabular}
\caption{Normalization factor for the LE point derived from a linear fit of the HE points, from Ref.~\protect\cite{Andivahis:1994rq}. The superscript "$*$" indicates the values that were directly derived from the ratio of the measured cross sections. }
\label{table1}
\end{center}
\end{table}
The results are reported in Fig.~\ref{Fig:W0} and compared to the ratio from polarization data. We may conclude that the results from Analysis II, III, and IV are consistent with the ratio extracted with the polarization method: a revision of the normalization factor brings the data into agreement. 

Moreover, at the light of all above, it is nonsense to use the FFs data from Ref.~\cite{Andivahis:1994rq} to probe the two-photon effect, as they were extracted under the hypothesis of linearity of the reduced cross section, i.e., correcting the first point  to be aligned.  As recalled above, a direct consequence of the presence of $2\gamma E$  would be the non-linearity of the Rosenbluth plot. Note that most $2\gamma E$ calculations were tuned precisely on this experiment.

In Ref. \cite{Pacetti:2015iqa} it was already noted that some unpolarized data, where radiative corrections were lower than $20\%$, indeed showed a deviation of the ratio $\mu^2 R^2$ from unity consistently with the polarization data.  A set of 64 data points, that include existing Rosenbluth data was reanalyzed in Ref. \cite{Pacetti:2016tqi}, in terms of FF ratio, confirming the compatibility among polarized and unpolarized elastic data. Further inspection shows that a somehow arbitrary renormalization of  subset of data was currently done. For example,  in Ref.~\cite{Litt197040}  one can read {\it  "changing the normalization of the small angle data from SLAC or DESY by $\pm 1.5\%$ with respect to the large angle data (Bonn})". This normalization increased the FFs ratio towards unity, according to the requirement that both electric and magnetic FFs follow a dipole dependence, as driven by pQCD scaling laws. The results showed consistency with the hypothesis $\mu^2 R^2\simeq 1$ at large $Q^2$, as expected at that time. A complete discussion and data basis of unpolarized and polarized measurements can be found in Ref.~\cite{Pacetti:2015iqa}.  
Among the available data from Rosenbluth separation, three sets~\cite{Walker:1993vj,Litt197040,Qattan:2004ht} show a particular behavior,  giving a value of the ratio that exceeds unity and  grows with $Q^2$. For these experiments it was noted in Ref.~\cite{TomasiGustafsson:2006pa} that radiative corrections and/or correlations are especially large. The data from Ref.~\cite{Qattan:2004ht} were extracted detecting the proton instead of the electron. Besides the above mentioned corrections, at large $Q^2$ the contamination of the elastic peak by the inelastic $e+p\to e+p+\pi^0$ reaction has to be carefully subtracted~\cite{Puckett:2011xg}. 

Let us note that for Refs.~\cite{Walker:1993vj,Litt197040}, $G_M^2$ extracted from the present analysis is systematically lower as compared to other data and global fits,  showing that these measurements may be affected by some systematic error probably due to normalization issues, whereas the results of Ref.~\cite{Qattan:2004ht} agree very well with the  standard parametrization of the magnetic contribution.

Concerning, in general, the elastic $ep$ cross section, several early experiments pointed out a deviation of the elastic cross section from the $(1/Q^2)^2$ behavior.  Quoting a  presentation of the data at the highest available transferred momenta, from Nobel prize R. Taylor: "There appears to be definite evidence in the data for a significant deviation from the dipole fit"~\cite{Taylor:1967qv}.  Radiative corrections were also already quoted as an issue to be treated with particular attention. 

\begin{figure}
  \begin{minipage}[c]{0.5\textwidth}
     \includegraphics[width=\textwidth, trim = 4mm 2mm 5mm 3mm, clip ]{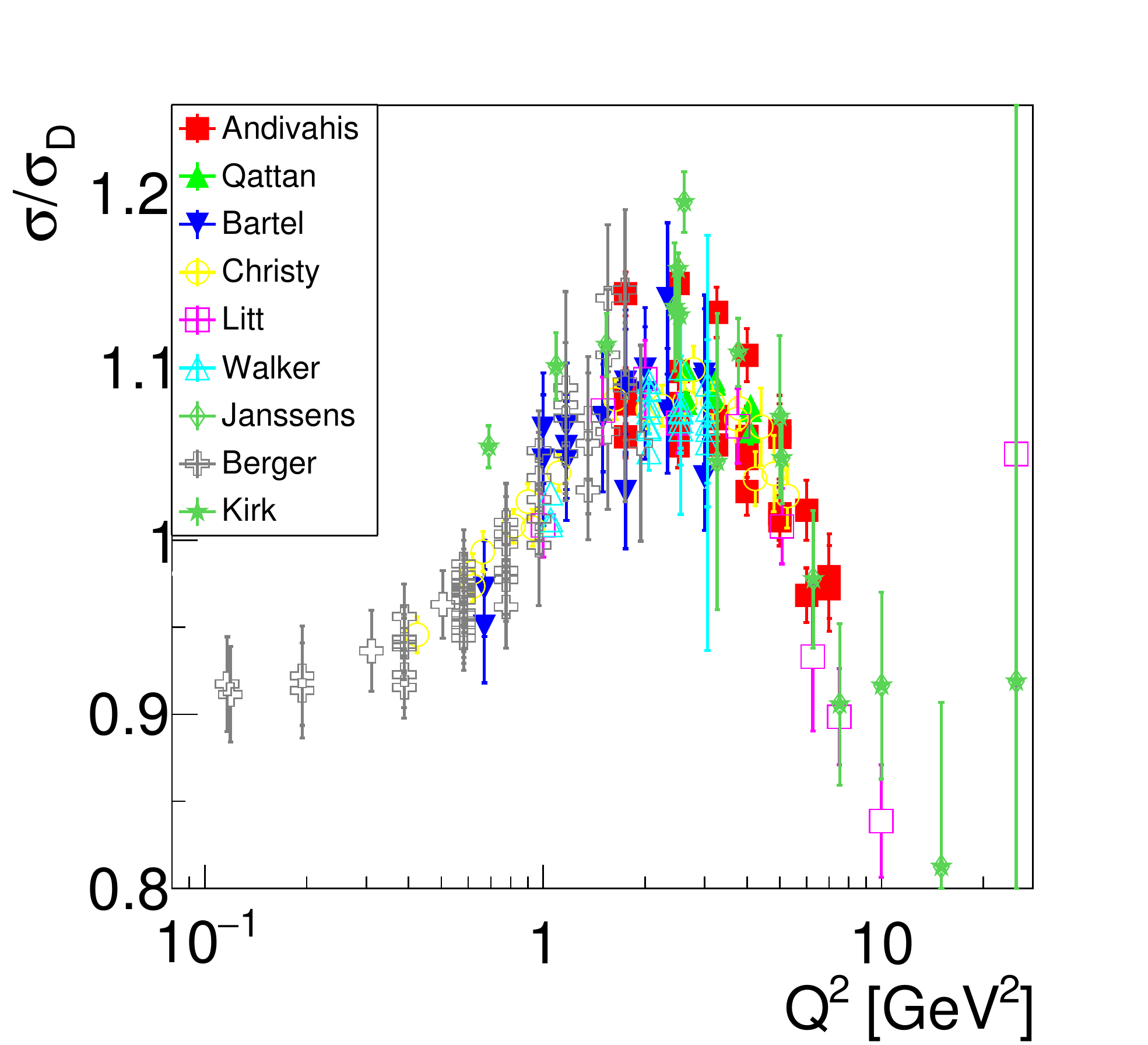}
      \end{minipage}   
    \begin{minipage}[c]{0.5\textwidth}
\caption{$\sigma/\sigma_D$  as a function of $Q^2$ for different experiments. 
 }
\label{Fig:dipole}
\end{minipage}
\end{figure}

The dipole normalized cross section
\be 
\frac{\sigma}{\sigma_D}=\frac{\sigma_{\rm red}^{\rm exp}}{ G_D^2(\epsilon/\mu^2 +\tau)}\,, 
\nonumber\ee
being $\sigma_{\rm red}^{\rm exp}$ the measured reduced cross section, is reported in Fig.~\ref{Fig:dipole} as a function of $Q^2$, regardless of the value of $\epsilon$. The $Q^2$ coordinates for the data  from a Rosenbluth separation for different $\epsilon$  are seen as vertically quasi-aligned symbols. Note that if these points form a cluster with overlapping error bars, it means that they are compatible with the relation $G_E\simeq G_M/\mu\simeq G_D$. If points are not overlapping, then FFs do not follow a dipole behavior. Concerning the data of Ref.~\cite{Andivahis:1994rq}, let us note that the dispersion at fixed $Q^2$ is not larger than the systematics from different sets. 

In general, and particularly at large $Q^2$, one can see that the dipole fit is not a good representation of the data. The deviation at large $Q^2$ reaches 20-30\% on the cross section and has to be attributed mainly to the magnetic term. This is very puzzling, as it is expected that the magnetic FF would follow quark counting rules, compatible with the $Q^2$ dipole dependence.  This mean that not only the electric FF but also the magnetic one differ from dipole, without compensation.

\section{Conclusions}
We have discussed the origin of the discrepancy among FFs derived from unpolarized elastic $ep$ scattering with the Rosenbluth separation and from the Akhiezer-Rekalo polarization method. 

We have proposed a reanalysis of the Rosenbluth data in terms of the squared FF ratio $R^2$ instead that of the extraction of the individual FFs, similarly to what has been done in the time-like region. In such a region, this procedure is more convenient because of the scarce statistics. In the present case it allows to consider $R^2$ as a parameter, directly extracted, avoiding the correlations between $G_E^2$, that is small and affected by large error bars, and $G_M^2$. The parameter $G_M^2$ here includes the eventual systematics and global normalization problems.

In general, the discrepancy between unpolarized and polarized experiments is not evident for the older experiments. Besides the large errors, most of them show indeed a decrease of the ratio, already reported in the literature. Up to 3-4 GeV$^2$ in some cases, the difference may be resolved by a proper calculation of radiative corrections. 

We point out inconsistencies in the claim of the presence of two-photon contributions.  The attempts to extract FFs  as real quantities, function of one variable, $Q^2$, in the presence of $2\gamma E$, is erroneous by principle.  In presence of two-photon effects one can not extract nucleon FFs from the unpolarized cross section.  The matrix element contains three amplitudes of complex nature, functions of two kinematical variables instead than two real functions of $Q^2$ only. 

Correcting the unpolarized  the cross section by an assumed two photon effect and re-extracting FFs, is erroneous, as it  integrates the conceptual and operative contradiction of merging the Born approximation and the two-photon effects. In all these analyses the FF extraction is based on the dominance of the $1\gamma E$ mechanism. Advocating a large contribution of the $1\gamma-2\gamma $  interference, would invalidate the definition of FF itself, as real function of the single variable $Q^2$. We do not enter here in the comparison and the merit of the existing model-dependent $2\gamma E$ calculations. 

{\it Does  the discrepancy between the unpolarized and polarized FF ratio experiments really exist?}

\noindent Following the recent work of Ref. \cite{Pacetti:2015iqa} a problem of renormalization of the low $\epsilon$ data in the previous data, in particular in  Ref. \cite{Andivahis:1994rq}, was pointed out. The discrepancy would remain only for the data from Ref. \cite{Qattan:2004ht}. For these data, the applied radiative corrections are not available, and a $\simeq$100\% correlation of the slope and the intercept (the parameters of the Rosenbluth fit) was pointed out in Ref.~\cite{TomasiGustafsson:2006pa}.

{\it Is the $2\gamma E$ contribution sizable? }

\noindent The analysis of the present and all data does not show an evident effect increasing with $Q^2$, beyond the expectation from $\alpha$-counting. Moreover, no theoretical strong argument has been put forward and confirmed by recent and old experiments to justify a large  $2\gamma E$ contribution.



\end{document}